\documentclass[usegraphicx,usenatbib,onecolumn]{mn2e}
\usepackage{bm}

\begin{document}
\title[The S-Z effect and Faraday rotation of galaxy groups]
{The Sunyaev-Zel'dovich effect and Faraday rotation contributions of galaxy groups to the
CMB angular power spectrum
 }

\author[Tashiro, H. et al.]
{Hiroyuki Tashiro$^1$, Joseph Silk$^2$, Mathieu Langer$^1$ and Naoshi Sugiyama$^{3,}$$^4$ \\
$^1$ Institut d'Astrophysique Spatiale (IAS), B\^atiment 121, F-91405, Orsay, France;\\
 Universit\'e Paris-Sud 11 and CNRS (UMR 8617)\\
$^2$ Astrophysics, Denys Wilkinson Building, The University of Oxford, Keble Road, Oxford OX1 3HR, UK\\
$^3$ Department of Physics and Astrophysics, Nagoya
University, Chikusa-ku, Nagoya,  464-8602, Japan \\
$^4$Institute for Physics and Mathematics of the Universe, 
University of Tokyo,\\ 
 5-1-5 Kashiwa-no-Ha, Kashiwa,
Chiba, 277-8582, Japan}

\date{\today}

\maketitle

\begin{abstract}

The S-Z effect and Faraday rotation from halos 
are examined over a  wide mass range, an including gas condensation and magnetic field evolution. Contributions to the
CMB angular power spectrum  are evaluated for galaxy clusters, galaxy groups and galaxies.
Smaller mass halos are found to play a more important role than  massive halos
for the $B$-mode polarisation associated with the S-Z CMB anisotropies.
The $B$-modes from Faraday rotation
dominate the secondary $B$-modes caused by gravitational
lensing at $\ell > 3000$.  
Measurement of $B$-mode polarisation in combination with the S-Z power spectrum
can potentially provide important  constraints on 
intracluster magnetic field and gas evolution at early epochs.

\end{abstract}


\maketitle

\section{Introduction}

The baryonic components in galaxy clusters, galaxy groups and galaxies  vary  widely in their evolution.
Some fraction of the baryons remains gaseous at the virial temperature 
and is threaded by microgauss magnetic fields, while a fraction condenses into the disk and forms stars. 
Therefore, 
knowing the temperature, density and the condensation rate of  the baryons 
is an important clue for revealing the history of star formation, galaxy formation
and even the magnetic field evolution and the reionisation process of the Universe.  

The Sunyaev-Zel'dovich effect (S-Z effect) is an important tool for the investigation
of baryon properties in halos \citep{sunyaev-zeldovich-1972}.
The S-Z effect is the scattering of cosmic microwave background (CMB) photons by the hot electron gas 
in the gravitational potential wells of halos.
As a result, the CMB frequency spectrum is distorted from the blackbody spectrum.
This distortion depends on the temperature and number density of the electron gas.
In combination with X-ray observations,
measuring the S-Z effect from galaxy clusters at each redshift gives 
the evolution of physical properties of the baryons \citep{morandi-ettori-2007}. 
Moreover, the temperature and the number density of the electron gas
are related to the masses of halos, so that
the number counts of detectable S-Z signals for different observational thresholds
can provide the mass distribution of galaxy clusters.
The distribution is sensitive to
the cosmological parameters and to the 
non-Gaussianity of primordial density fluctuations.
Therefore, the number counts of S-Z galaxy clusters can also give a constraint
on the cosmological model
\citep{bartlemann-sz-2001,moscardini-bartlemann-2002, schafer-pfmmer-2006, sadeh-rephaeli-2007}.
Many future CMB surveys will make important contributions, and in some cases are even  dedicated, to the measurement of S-Z-selected  galaxy clusters
(e.g. {\sc Planck}\footnote{http://www.rssd.esa.int/index.php?project=Planck},
SPT\footnote{http://pole.uchicago.edu/}, 
ACT\footnote{http://wwwphy.princeton.edu/act}).

The S-Z effect from unresolved S-Z clusters 
is observed as CMB temperature anisotropies.
These anisotropies are one of the major secondary anisotropies
                                                                            fand dominant components for the high $\ell$ modes ($\ell >2000$)
in the angular temperature power spectrum.
The amplitude strongly depends on
the amplitude of matter density fluctuations, $\sigma_8$.
Small angular-scale CMB anisotropy experiments,
BIMA \citep{dawson-bima}, CBI \citep{mason-cbi}
and ACBAR \citep{kuo-acbar},
have measured CMB anisotropies at high $\ell$ modes 
and detected the excess signal expected to be due to the S-Z effect.
This excess corresponds to the S-Z effect with $\sigma_8 \sim 1.0$ \citep{bond-szsigma, 
douspis-aghanim-2006}.
However, this high value is in conflict with the 5 year WMAP result, $\sigma_8 \sim 0.8$, 
which is obtained from large-scale temperature anisotropies \citep{wmap-spergel-07}.
Note that this conflict is lifted when a possible contribution from unresolved 
point sources is taken into account \citep{douspis-aghanim-2006}. 
In order to check whether the excess indeed is due to  the effect of large $\sigma_8$,
we need detailed observational data on the S-Z number counts by future observations.  

In this paper, we re-examine the number counts of S-Z halos 
and S-Z CMB anisotropies.
Our aim is to clarify which
halos with different masses contribute to any particular $\ell$ range.
In particular, we focus on 
the S-Z effect of galaxy groups and galaxies ($M<10^{14} M_\odot$). 
In those halos,
the condensation of baryonic gas is effective. 
Clearly then,
the S-Z effect is expected to depend on
the evolution of gas condensation.  

We also study Faraday rotation by magnetic fields in galaxy clusters, galaxy groups and galaxies.
Many observations suggest that galaxy clusters and galaxies have magnetic fields whose
amplitude is typically of the order of $1$--$10$ $\mu$Gauss.  
The coherence length of such magnetic fields can be as
large as cluster virial scales,
and structured magnetic fields have been found even on larger scales.
\citep{kim-kronberg-1989}.
Besides, magnetic fields have been measured with the same amplitudes
in high redshift objects at $z>2$ \citep{athreya-kapahi-1998}.
However, the evolution of these fields has hitherto been unclear
and is the one of the open questions in modern cosmology.
In order to answer this question,
we need statistical discussions of magnetic fields.
In this paper, we discuss the potential of Faraday rotation 
as a probe of magnetic fields of galaxy clusters, galaxy groups and galactic halos
by studying two observational methods.
One is the number counts of Faraday rotation for S-Z halos.
We calculate the number counts,
and discuss the sensitivity of properties and evolution of magnetic fields
in halo objects.

The other is the CMB $B$-mode polarisation observation.  
Faraday rotation distorts the CMB polarisation fields like 
gravitational lensing, so that it induces the $B$-mode polarisation from
the $E$-mode.
The angular power spectrum due to this effect in halos
was studied by \citet{takada-ohno} and \citet{tashiro-faraday}.
In particular, \citet{tashiro-faraday} pointed out
that the angular power spectrum is sensitive to the magnetic field evolution
and magnetic fields in galaxies can make a more dominant contribution 
than those in galaxy clusters.
However, their adopted magnetic fields are simple toy models.
Therefore, for a more detailed discussion, 
we study the CMB $B$-mode polarisation 
due to magnetic fields in halo objects 
based on observations and numerical simulations.

The paper is organised as follows.  In
Sec.~II, 
we discuss the profiles of electron density and magnetic fields
that we use, motivated by observations and numerical simulations.
We calculate the number counts of halos for the S-Z effect and Faraday rotation
in Sec.~III.
We discuss the impact of gas condensation and magnetic field evolution
on the number counts.
In Sec.~IV, 
we compute the power spectrum 
of CMB temperature anisotropies by the S-Z effect and
$B$-mode polarisation due to Faraday rotation.
Sec.~V is devoted to discussion and summary.  
Throughout
the paper, we use the following cosmological parameters:
$h=0.7 \ (H_0=h \times 100 {\rm km/s / Mpc})$, $T_0 = 2.725$K, 
$\Omega _{\rm B} =0.044$ and $ \Omega_{\rm M} =0.27$ and $\sigma_8 =0.8$.

\section{halo models}\label{sec:profile}

\subsection{Electron density and temperature profiles}

For the electron density and temperature profiles in halos, 
we use the results of \citet{komatsu-seljak},
which are based on the NFW dark matter density profile \citep{navarro-frenk}. 
The NFW dark matter density profile is given by 
\begin{equation}
\rho_{\rm DM}(x) = \frac{\rho_{\rm s}}{x(1+x)^2}.
\label{eq:NFWprofiles}
\end{equation}
Here, $x \equiv r/r_{\rm s}$ where $r_{\rm s}$ is a scale radius, and
$\rho_{\rm s}$ is a scale density. The scale radius
$r_{\rm s}$ is related to the virial radius by 
the concentration parameter $c$
\begin{equation}
{r_{\rm s}(M,z)}={r_{\rm vir}(M,z) \over c(M,z)}.
\label{eq:scaleradius}
\end{equation}
In the following, we adopt
the concentration parameter of \citet{komatsu-seljak} where
\begin{equation}
c(M, z) \approx \frac{10}{1+z}\left[\frac{M}{M_*(0)}\right]^{-0.2}.
\label{eq:concentrait}
\end{equation}
Here $M_*(0)$ is a solution to $\sigma(M)=\delta_c$ at the redshift $z=0$
where $\sigma$ is the variance smoothed with a
top-hat filter of  scale $R= ( 3M/4 \pi \bar \rho)^{1/3}$.

In order to obtain the profiles of the electron number density $n_e$ 
and temperature $T_e$,
\citet{komatsu-seljak}
considered three assumptions: (i) the electron gas is in hydrostatic
equilibrium in the dark matter potential, (ii) the electron gas density follows the
dark matter density in the outer parts of the halo, and (iii) the
equation of state of the electron gas is polytropic, $P_{e}\propto \rho_{e}^{\gamma}$ 
where $P_{e}$, $\rho_{e}$ and ${\gamma}$
are the electron gas pressure, the gas density and the polytropic index, respectively. 
Under 
these assumptions, %with $n_e \propto \rho_{e} /\mum_p$, 
the electron number density and temperature profiles 
are simply given by
\begin{equation}
n_e = n_{e \rm c} F(x),
\end{equation}
\begin{equation}
T_e = T_{e \rm c} F^{\gamma -1}(x).
\end{equation}
Here, the dimensionless profile $F(x)$ is written as
\begin{equation}
F(x) = \left \{ 1-A \left[ 1- {\ln (1+c) \over x}\right]
\right \}^{1/(\gamma -1)},
\end{equation}
where the coefficient $A$ is given by
\begin{equation}
A\equiv3\eta^{-1}_{\rm c}\frac{\gamma-1}{\gamma}
\left[\frac{\ln(1+c)}c-\frac1{1+c}\right]^{-1}.
\label{eq:Bcoefficient}  
\end{equation}
For $\gamma$ and $\eta_{\rm c}$,
\citet{komatsu-seljak} provided the following useful fitting formulae for
$\gamma$ and $\eta_{\rm c}$,
\begin{equation}
\gamma= 1.137 + 8.94\times 10^{-2}\ln(c/5) - 3.68\times 10^{-3}(c-5),
\label{eq:gammafitting}
\end{equation}
\begin{equation}
\eta_{\rm c}= 2.235 + 0.202(c-5) - 1.16\times 10^{-3}\left(c-5\right)^2.
\label{eq:etafitting}
\end{equation}
The central electron density $n_{e \rm c}$ and the 
temperature $T_{e \rm c}$ are obtained as
\begin{equation}
n_{e\rm c}
=3.01 
\left( {M \over 10^{14} M_\odot } \right)
\left( {r_{\rm vir} \over 1~ {\rm Mpc}  } \right)^{-3}
f_{\rm g}
\frac{c}{(1+c)^2}
\left[\ln(1+c)-\frac{c}{1+c}\right]^{-1}
\left[{c \over c - F [c-\ln(1+c)] }\right]^{1/(\gamma-1)}
~{\rm cm}^{-3}     ,
\label{eq:rhog0}
\end{equation}
\begin{equation}
T_{e\rm c}
=0.88  
~\eta_{\rm c} 
\left[ M / (10^{14} h^{-1} M_{\odot}) \over 
r_{\rm vir} / (1 h^{-1} {\rm Mpc}) \right]
~{\rm keV},
\label{eq:rhog0}
\end{equation}
where the gas fraction $f_{\rm g}$ represents the fraction of the gas that remains
in the dark matter halo.
The upper limit of this function is the 
background fraction of the baryon to the dark matter density $\Omega_{\rm B}/ \Omega_{\rm DM}$.

\citet{gonzalez-zaritsky-2007} gave the fitting formula of the present day gas fraction 
as $\log  f_{\rm g0}(M) =  -3.87  + 0.2 \log M$.
In the early universe, gas condensation is not well advanced 
and the gas fraction can be assumed to be almost unity.
As the universe evolves, gas condensation proceeds and star formation occurs 
in the condensed disk.
Therefore, we assume that the evolution of the gas fraction is related to star formation by
\begin{equation}
f_{\rm g}(z,M) = 1 - {g(z)  \over g(0)}[1-f_{\rm g0}(M)],
\end{equation}
\begin{equation}
g(z) = \int ^z d z' ~f_*  ~dV/ dz'   .  
\end{equation}
Here, $f_*$ is the star
formation rate per comoving volume. 
We use the data in \citet{bouwens-illingrorth-2007} in order to obtain $f_*$.
We show $f_{\rm g}(z,M) $ in Fig.~\ref{fig:gasfraction}. 
The gas condensation starts around $z=5$ in halos with mass lower than $10^{15} M_\odot$.
However, in halos with mass larger than $10^{15} M_\odot$,
the gas condensation is not effective 
and $f_{\rm g}$ almost equals  the 
background ratio of baryonic to dark matter density $\Omega_{\rm B}/ \Omega_{\rm DM}$.

\subsection{Magnetic field distribution}

The magnetic field profile in halos
is not known in any detail.
Observations by \citet{murgia-govoni}
suggest that the distribution of
magnetic fields in galaxies and galaxy clusters is such that their strength
decreases outwards.  
Accordingly, 
we consider two assumptions for magnetic fields in halos.
The first one is that magnetic fields are frozen into the matter
and therefore follow the gas density profile.
The second is that the fields reach equipartition with thermal pressure
in the central part in the halo.
%{\bf
%The magnetic field strength at the centre of a halo has not been measured,
%although there are many observations of the measurement of  
%the magnetic field strength in a halo except the centre.
%Therefore, we utilize the equipartition assumption although it is not especially motivated 
%by theorety.
%}
Under these assumptions, the present day magnetic field profile is written as
\begin{equation}
B(x) \propto B_c F ^{2/3} (x) ,
\end{equation}
\begin{equation}
B_c = 2.0 \left ({M \over 10 ^{14} M_\odot} \right)
\left ({r_{\rm vir} \over 1 h^{-1} {\rm Mpc}} \right)^{-2/3} 
\mu {\rm Gauss},
\end{equation}
where $B_c$ is the magnetic field strength at the centre of the halo with mass $M$,
and is proportional to $ M^{1/3}$.

The magnetic field evolution, especially in galaxy clusters, is unknown.
Here, we adopt two cases for magnetic field evolution scenario.
The first case is motivated by theoretical considerations,
and the second case is based on numerical simulations.

\subsubsection*{\underline{CASE I}}

In the first case,
we assume that the magnetic fields strength depends on 
the star formation  rate via  the evolution of the gas fraction.
The magnetic field strength at redshift $z$ is then given by
\begin{equation}
B(z)= \sqrt{{g(z)  \over g(0)}} B_0,
\end{equation}
where $B_0$ is the magnetic field strength at present.
This assumption is based on the theoretical idea that 
stars play an important role in the generation of magnetic fields.
The seed magnetic fields are produced in stars and 
then spread over the halo 
by supernovae, AGN jets and galactic winds.
They evolve rapidly over a dynamical time scale 
by the $\alpha$-$\Omega$ dynamo in disks and small scale fluctuation dynamo in clusters
\citep{brandenburg-subramania-2005}.
Finally, they reach micro-Gauss magnetic field strengths. 
In this case, the magnetic field strength is reduced to half of its present day value
at $z \sim 3$.

\subsubsection*{\underline{CASE II}}

Case II is based on the numerical simulations of \citet{dolag-bartelmann-2002}.
They showed that the evolution of magnetic fields in galaxy clusters goes like 
$B(z) \propto e^{-2.5 z}$.
In this case 
we assume that the fields  evolve as 
\begin{equation}
B(z)=\left\{
  \begin{array}{cc}
     e^{-2.5 z} B_0,  &  M \ge M_*,  \\
       \sqrt{g(z)  /g(0)} B_0,    &  M < M_*,  \\
  \end{array}
\right.
\end{equation}
where $M_*$ is the mass of the halo
whose dynamical time scale equals
the cooling time scale.
In halos with mass smaller than $M_*$,
the cooling time scale is shorter than the dynamical time scale.
Therefore, such halos are expected to become star-forming galaxies.
On the contrary,
halos with mass larger than $M_*$, whose cooling time scales
are larger than the dynamical time scale,
end up as galaxy clusters or galaxy groups.

\begin{figure}
  \begin{center}
    \includegraphics[keepaspectratio=true,height=50mm]{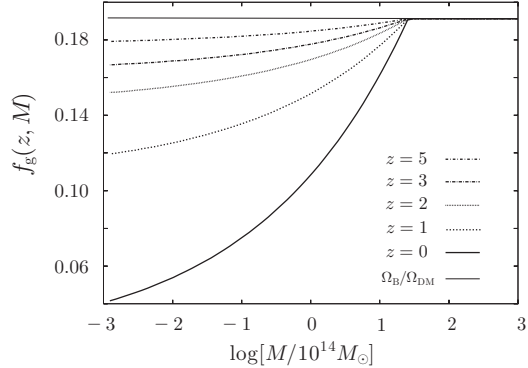}
  \end{center}
  \caption{The gas fraction as a function of dark matter halo mass. 
  The lines show the gas fractions in different redshifts: 
  $z=0$, $z=1$, $z=2$, $z=3$ and $z=5$ from bottom to top.
  For comparison, we plot 
  $\Omega_{\rm B} /\Omega_{\rm DM}$ as a thin line.} 
   \label{fig:gasfraction}
\end{figure}

\section{halo number counts}

Galaxy clusters, galaxy groups and galaxies are good probes of the large scale
structure of the Universe.
For the S-Z effect, we can easily resolve a halo from the CMB sky map. 
The resolved number of halos
for different masses and redshifts
depends not only on the cosmology 
but also on the baryon physical properties. 
In particular, gas condensation has a serious effect
on the number counts.
In this section, we calculate the number counts,
modeling gas condensation with a recipe based on recent observations.
Moreover, we calculate the number counts for Faraday rotation.
Faraday rotation in 
combination with S-Z effect is expected to provide
a new  constraint on magnetic fields in halos.

\subsection{S-Z halo number counts}

The change of the CMB intensity due to the S-Z effect can be written 
in terms of the Compton $y$-parameter as
\begin{equation}
{\Delta I \over I_0} = Q(p) y,
\label{eq:changeintensity}
\end{equation}
where $p = h_p \nu /k_B T_{\rm CMB}$ and
$I_0 = (2 h_p /c^2)(k_B T_{\rm CMB})^3/h_p^3$
with the Boltzmann constant $k_{\rm B}$
and the Planck constant $h_p$.
In Eq.~(\ref{eq:changeintensity}), $Q(p)$ is given by
\begin{equation}
Q(p) = {p^4 e^p \over (e^p -1 )^2} 
\left[ {p \over \tanh (p/2) }-4 \right].
\end{equation}
The Compton $y$-parameter is given by
\begin{equation}
y = \int dl~ {\sigma_{\rm T} \over m_e} ~ n_e k_{\rm B}T_e ,
\end{equation}
where 
%$T_e$ is the temperature of the gas in the cluster, $m_e$ is the electron mass, 
%$n_e$ is the electron number density,
$\sigma_{\rm T}$ is the Thomson scattering cross section, and the
integral is calculated along the line-of-sight in the cluster.

In order to discuss the S-Z signal  
from unresolved halo objects independently of frequency,
it is convenient to define the quantity $Y$ 
which is the value obtained by the integration of the $y$-parameter
over the surface area of the halos,
\begin{equation}
Y= {1 \over D_A^2(z)} \int d A ~ y(x) ,
\label{eq:def-largeYpara}
\end{equation}
where $D_A(z)$ is the angular diameter distance to the halo at redshift $z$
and $Y$ has a unit of a solid angle.
For example, for the {\sc Planck} satellite, the observable S-Z effect limit $Y_{\rm lim}$ corresponds 
to %Y_{\rm lim}=
$3 \times 10 ^{-4} {\rm arcmin}^2$ 
and can be derived from the optimal antenna temperature \citep{bartlemann-sz-2001}.

The temperature profile in halos, which is obtained in Sec.~\ref{sec:profile}, 
is almost the isothermal profile.
Therefore, it is valid to apply the isothermal assumption to Eq.~(\ref{eq:def-largeYpara})
and we obtain
\begin{equation}
Y={f_{\rm g} \over D_A^2(z) }{1+X \over 2}  
{M \over m_p} \sigma_T {k T_e \over m_e c^2},
\end{equation}
where $X$ is the hydrogen ratio and is $X=0.76$.

In order to calculate the number counts,
we need to know the mass function of halos, ${dn(M,z)/dM}$.
We adopt the fitting formula given by \citet{seth-tormen},
\begin{equation}
{dn(M,z) \over dM} = {\bar \rho \over M} \left(1+2^{-p} {\Gamma(1/2-p)\over  \sqrt \pi} \right)
(1+(q \nu)^{-p})\left({q \nu \over 2 \pi}\right)^{1/2}
\exp \left(- {q \nu \over 2}\right){d \nu \over \nu},
\label{eq:fitting-formula}
\end{equation}
where $\nu = \left( \delta_c \over \sigma(M,z) \right)^2$, $\delta_c$ is the
critical over-density and $\sigma$ is the variance smoothed with a
top-hat filter of a scale $R= ( 3M/4 \pi \bar \rho)^{1/3}$ and   
we take $p = 0.3$, $q=0.75$ \citep{cooray-sheth-02}.

\begin{figure}
  \begin{tabular}{cc}
   \begin{minipage}{0.5\textwidth}
  \begin{center}
    \includegraphics[keepaspectratio=true,height=50mm]{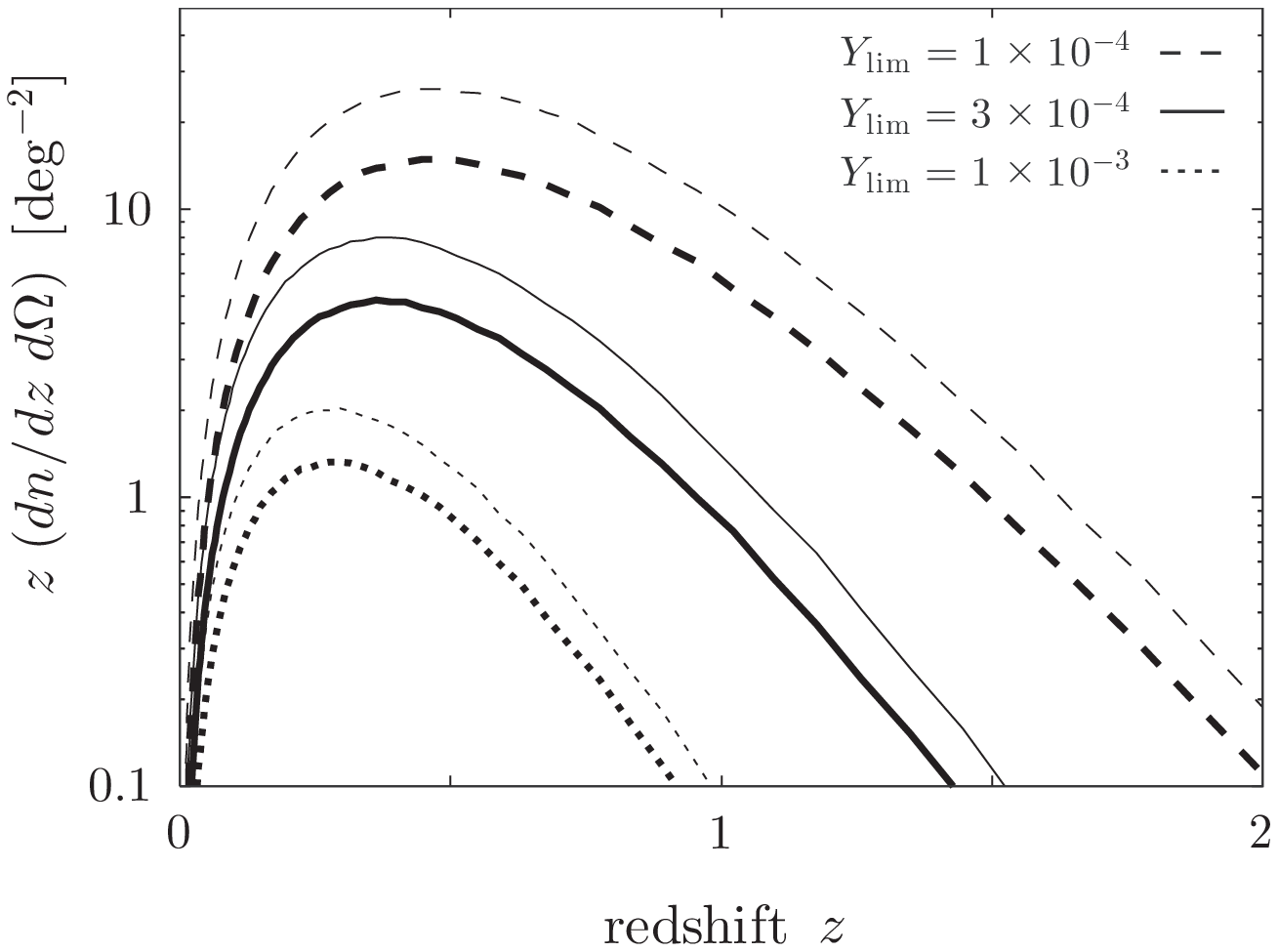}
  \end{center}
  \end{minipage}
   \begin{minipage}{0.5\textwidth}
  \begin{center}
    \includegraphics[keepaspectratio=true,height=50mm]{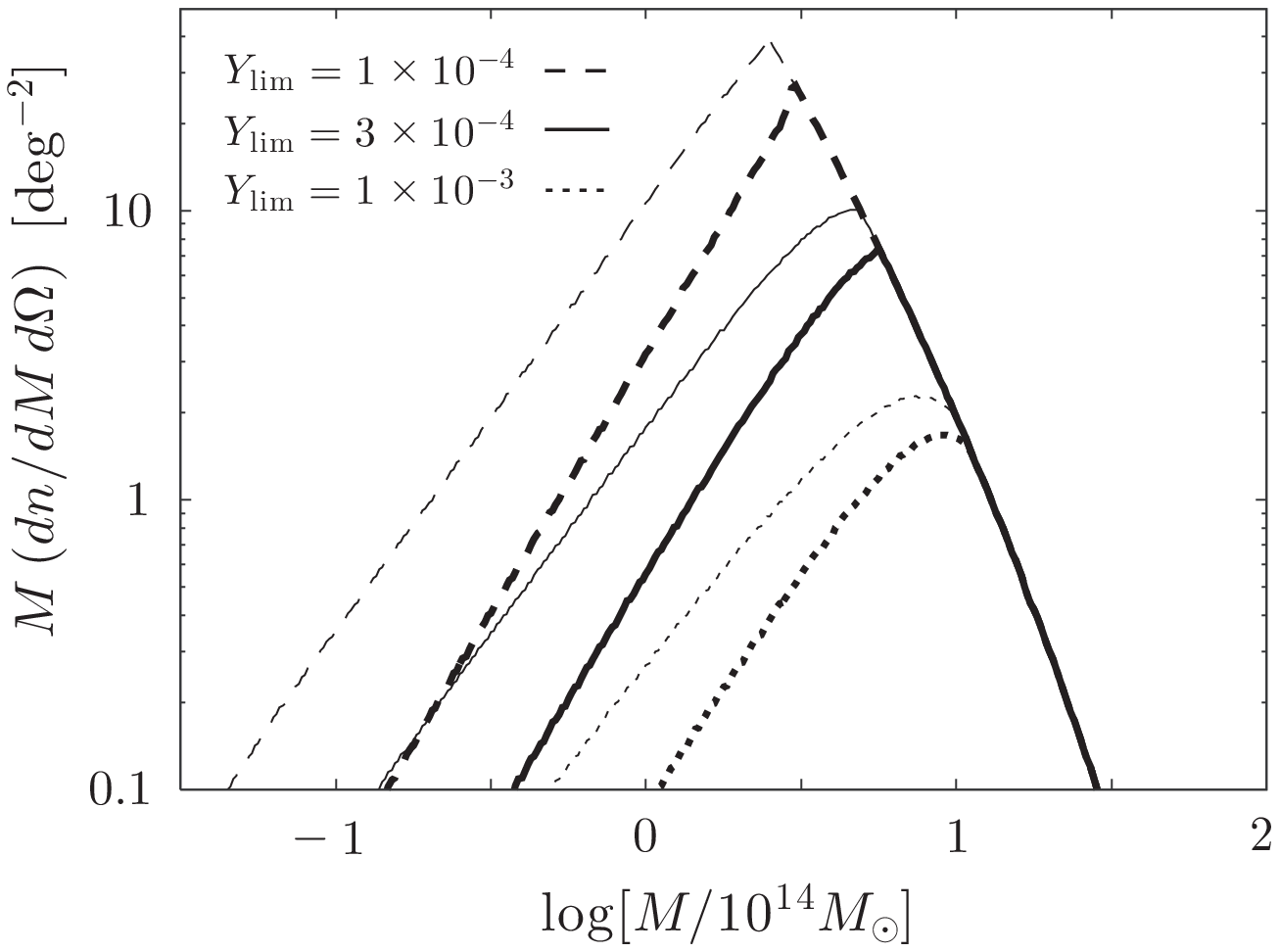}
  \end{center}
   \end{minipage}
  \end{tabular}
  \caption{
  The number counts of S-Z halos. 
  In both panels, the dashed, solid and dotted lines represent 
  the number counts for $Y_{\rm lim}=1.0\times 10^{-4}$,
  $3.0\times 10^{-4}$ and $1.0\times 10^{-3}$, respectively. 
  For comparison, we put the number counts 
  without gas condensation
  as thin lines.
  In the left panel, we show the number counts of S-Z halos as 
  a function of redshift.
  In the right panel, we plot the number counts of S-Z halos as 
  a function of halo mass.
  }
  \label{fig:ypara_red}
\end{figure}

We plot the number counts of S-Z halos for a given $Y_{\rm lim}$ 
as a function of redshift in the left panel of Fig.~\ref{fig:ypara_red}. 
For comparison, we plot the number count distribution
without gas condensation ($f_g  = \Omega _{\rm B} /\Omega_{\rm DM}$ 
on all mass scales) as thin lines.
This figure shows that
most of the contribution comes from redshifts lower than $0.5$.
The difference between the cases with and without gas condensation
becomes a little bit larger at low redshifts than at high redshifts.
We find that 
the peak position of the number count does not depend on gas condensation,
and
the number count in the case with gas condensation at 
redshifts lower than 0.5
is about half of that without gas condensation.

The right panel in Fig.~\ref{fig:ypara_red} shows 
the number counts of S-Z halos for a given $Y_{\rm lim}$ 
as a function of mass.
In the spherical model, the electron temperature
for given mass and redshift is proportional
to $M^{2/3} (1+z)$.
Since halos with larger masses produce stronger signals,
they are easy to detect.
Therefore, the number counts on the right side of the peak 
do not depend on $Y_{\rm lim}$ and are determined only 
by the mass function of halos.
This is why all plots for different $Y_{\rm lim}$ lie on the same line.

The electron temperature is proportional to $z+1$.
Besides, the angular diameter distance increases until about $z \sim 2$,
and declines for higher redshifts.
Accordingly, the S-Z signal $Y$ reaches a minimum around $z \sim 1$.
At this redshift, the signal becomes lower than $Y_{\rm lim}$ for halos 
whose mass is smaller than a certain corresponding mass $M_{\rm lim}$.
The peak position is determined by this mass, and its 
position depends strongly on the gas fraction.
The sharpness of the peak strongly depends on
the derivative of the mass function at $Y_{\rm lim}$.
For high $Y_{\rm lim}$, $M_{\rm lim}$ is large 
so that the slope of mass function at this mass scale is steep.
On the contrary, a lower $Y_{\rm lim}$ corresponds
to a smaller $M_{\rm lim}$ where the mass function is almost flat.
%The number of such halos around $z \sim 1$ is large.
%because it is not the tail part but the middle part
%in the mass function of redshift. 
%Imposing $Y_{\rm lim}$ means 
%that large amount of halos around $z \sim 1$ are excluded.
%Therefore, the sharp edge is created.

The gas condensation has an effect on the number counts on small mass scales.
The smaller the halo mass, 
the larger is the deviation between the cases  with and without gas condensation. 
This is because gas condensation is important for  the small mass halos as shown 
in Fig.~\ref{fig:gasfraction}.

\subsection{Faraday rotation measurements}

Next, we investigate Faraday rotation in the S-Z halos.
The change of the polarised rotation angle by Faraday rotation is 
written as
\begin{equation}
\alpha = \lambda ^2 RM.
\end{equation}
Here, $\lambda$ is the wave-length 
and $RM$ is the rotation measure, which characterises the Faraday rotation,
given by
\begin{equation}
RM ={e^3 \over 2 \pi m_e ^2} \int dl~ n_e B  \hat \gamma \cdot \hat b.
\label{eq:faraday-formula}
\end{equation}
where $B$ is the magnetic field strength along the line of sight,
and $\hat \gamma$ and $\hat b$ are the orientations of the line of sight and magnetic fields.

We calculate the rotation measure
along the path that intersects the centre of the S-Z halos.
For simplicity,
we also assume that the coherence length of magnetic field is the virial radius 
and that the orientations are random.
This assumption yields $\langle | \hat \gamma \cdot \hat b | ^2 \rangle =1/3$.

\begin{figure}
  \begin{tabular}{cc}
   \begin{minipage}{0.5\textwidth}
  \begin{center}
    \includegraphics[keepaspectratio=true,height=50mm]{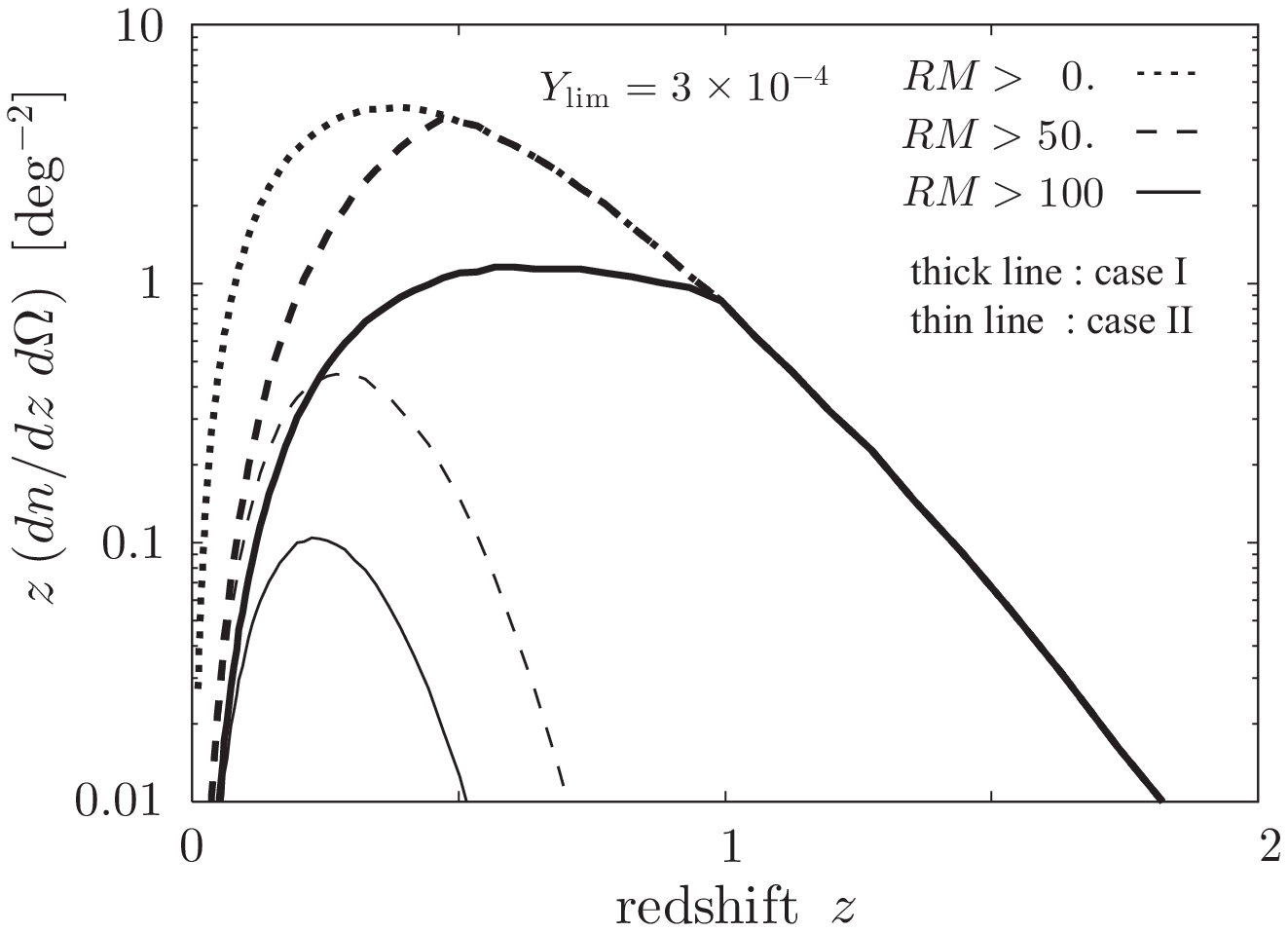}
  \end{center}
  \end{minipage}
   \begin{minipage}{0.5\textwidth}
  \begin{center}
    \includegraphics[keepaspectratio=true,height=50mm]{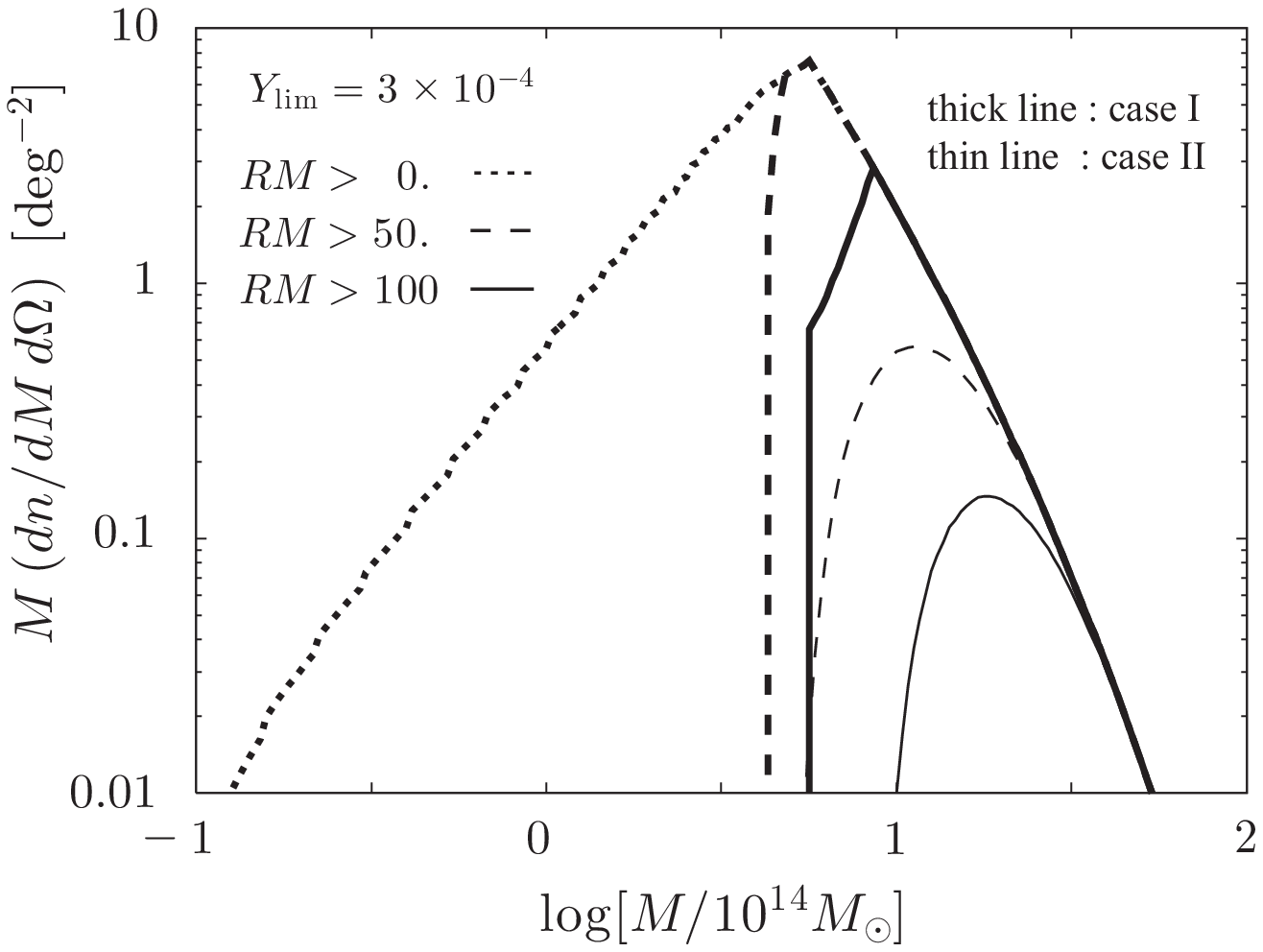}
  \end{center}
   \end{minipage}
  \end{tabular}
  \caption{ The number counts of S-Z halos with $Y_{\rm lim}=3.0\times 10^{-4}$ 
  for rotation measurements. In both panels,
  the number counts for $RM>0$, $RM>50$ and $RM>100$ are plotted as
  dotted, dashed and solid lines, respectively. 
  The thick lines are for case I and the thin lines are for case II.
  %{\bf
  %The total halo number per ${\rm deg}^2$ $dn /dm d \Omega$ is same 
  %in both case I and case II because the total halo number 
  %depends on the cosmological parameters. Therefore, 
  %the number for $RM > 0$ is also same in both cases, 
  %although the number distribution of $RM$ for case II is different.}
  In the left panel, we give the number counts as a function of redshift.
  In the right panel we plot the number counts as a function of halo mass. 
  }
  \label{fig:rot_red_y3}
\end{figure}

We plot the number distribution of the rotation measure
as a function of the redshift in the left panel of Fig.~\ref{fig:rot_red_y3}.
We have set $Y_{\rm lim} = 3 \times 10^{-4}$
and we show the number counts for  case I 
and case II of the magnetic field evolution.
As we increase the threshold of the rotation measurement, 
we fail to detect small mass halos, 
because the rotation measurement strongly depends on the halo mass.
This failure accounts for the lack of  detectable numbers of halos 
in low redshifts.
In case I, 	
the magnetic field evolution is very slow,
so that the halos with large mass, even at high redshifts,
produce a large rotation measurement.
Since S-Z halos at high redshifts have large masses, 
we can detect all of them in case I.
On the other hand,  case II has rapid magnetic field evolution at low redshift.
Therefore, since the amplitude of magnetic fields becomes very small at high redshifts, 
we cannot detect the signal from high redshift halos.
As a result, the distribution of the number counts over redshift in  case II
is suppressed.

The right panel in Fig~\ref{fig:rot_red_y3} shows the number count distribution
as a function of mass.
As in Fig.~\ref{fig:rot_red_y3},
we have set $Y_{\rm lim} = 3 \times 10^{-4}$
and we plot the number counts for  case I 
and case II of magnetic field evolution.
The threshold determines the detectable minimum mass of S-Z halos at each redshift.
Although the minimum mass at present does not depend on the evolution of magnetic fields,
as the redshift increases, a difference arises between cases I and II.
In  case II where the evolution is rapid,
since even large mass halos cannot produce  rotation measures above the threshold,
we can detect fewer halos than in  case I.
However, halos with sufficiently large masses, even at high redshift, 
can produce larger rotation measures.  
The peak of the number counts is located at this critical  minimum mass
and all S-Z halos with larger  masses are  detectable.
The peak position 
is sensitive to the threshold and the evolution of magnetic fields. 
This implies that
varying the threshold of the rotation measurement
can give us the number counts of S-Z halos with the
corresponding mass scale and a constraint
on magnetic field evolution.

\section{CMB anisotropies due to S-Z effect and Faraday rotation}

\subsection{Angular power spectrum of S-Z effect and Faraday rotation}

Halos with S-Z signals lower than $Y_{\rm lim}$
contribute to the diffuse temperature anisotropies and
cannot be individually removed from the CMB map. 
The computation of the angular power spectrum of S-Z temperature anisotropies is 
based on the halo formalism \citep{cole-kaiser, makino-suto, komatsu-kitayama},
\begin{equation}
C_\ell ^{SZ}= g_\nu^2 \int_0^{z_{\rm max}} dz \frac{dV}{dz}
\int_{M_{\rm min}}^{M_{\rm max}} dM \frac{dn(M,z)}{dM}
\left|y_\ell(M,z)\right|^2,
\label{eq:cly}
\end{equation}
where $g_\nu$ is the spectral function of the S-Z effect, which is 
$-2$ in the Rayleigh-Jeans limit.
In Eq.~(\ref{eq:cly}),
$y_\ell(M,z)$ is the 2D Fourier transform of the projected Compton
$y$-parameter and
is obtained from 
\begin{equation}
y_\ell= {4\pi r_{\rm s} \over \ell_{\rm s}^2}
\int_0^\infty dx x^2 y_{3d} (x)
{\sin(\ell x/\ell_{\rm s})  \over \ell x/\ell_{\rm s}},
\label{eq:yl}
\end{equation}
where $y_{3d}$ is the radial profile of Compton
$y$-parameter,
\begin{equation}
y_{3d} (x) =  {\sigma_{\rm T} \over m_e} ~ n_e(x) k_{\rm B}T_e (x),
\end{equation}
and $\ell_{\rm s}$ is the angular wavenumber corresponding to $r_{\rm s}$,
$\ell_{\rm s} = D_A / r_{\rm s}$.

Faraday rotation in halos produces the secondary $B$-mode polarisation 
from primary $E$-mode polarisation.
The produced $B$-mode angular power spectrum is given by \citep{takada-ohno, tashiro-faraday}
\begin{equation}
C^{ {\rm Fara} }_\ell= N_\ell^2 \sum_{\ell_1\ell_2}
N_{\ell_2}^2 K(\ell,\ell_1,\ell_2)^2 C^{E}_{\ell_2} C^\alpha_{\ell_1}
{(2\ell_1+1)(2\ell_2+1)\over 4\pi(2\ell+1)}\left(C^{\ell0}_{\ell_1 0 \ell_2 0} \right)^2 ,
\label{eq:B--mode-result}
\end{equation}
where $C^{c\gamma}_{a \alpha b \beta}$ are the Clebsch-Gordan coefficients,
$N_\ell = (2(\ell-2)!/(\ell+2)!)^{1/2}$ and 
\begin{equation}
K(\ell,\ell_1,\ell_2)\equiv -{1\over 2}\left(L^2 + L_1^2 + L_2^2 -2L_1L_2
-2L_1L +2L_1-2L_2 -2L\right),
\label{eq:k-factor}
\end{equation}
with $L=\ell(\ell+1)$, $L_1=\ell_1(\ell_1+1)$, and $L_2=\ell_2(\ell_2+1)$. 
In Eq.~(\ref{eq:B--mode-result}),
$C^{E}_\ell$ is the primary $E$-mode polarisation angular spectrum 
and $C^\alpha_{\ell}$ is the angular power spectrum of the rotation measurements.
The expression of $C^\alpha_{\ell}$ is 
\begin{equation}
C_\ell ^\alpha= \int_0^{z_{\rm dec}} dz \frac{dV}{dz}
\int_{M_{\rm min}}^{M_{\rm max}} dM
{dn(M,z) \over dM} \left|\alpha_\ell (M,z)\right|^2,
\label{eq:power-rotationmeasure}
\end{equation}
where
$\alpha_l(M,z)$ is the projected Fourier transform of the rotation
angle obtained in the small angle approximation,
\begin{equation}
\alpha_l = 2 \pi \int d \theta \theta \alpha(\theta,M, z) J_0(l \theta).
\label{eq:angularfourier}
\end{equation}
Here $\alpha(\theta,M, z)$ is the angular profile of the rotation
measurement induced by magnetic fields in a halo with mass $M$ at redshift $z$
and $\theta$ is the angular separation from the centre of a halo.

\subsection{Results}

\begin{figure}
  \begin{center}
    \includegraphics[keepaspectratio=true,height=50mm]{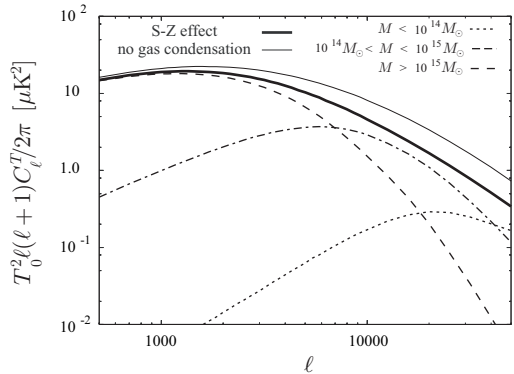}
  \end{center}
  \caption{Angular power spectra of the S-Z effect. The contributions
  from collapsed halo objects with $M < 10^{14} M_\odot$, 
  $10^{14} M_\odot <M < 10^{15} M_\odot$ and $M> 10^{15} M_\odot$ are represented 
  as dotted, dotted-dashed, dashed lines, respectively.
  For comparison, we plot the spectrum of the S-Z effect without gas condensation.} 
   \label{fig:Ycl_evo_comp}
\end{figure}

First,  we plot the S-Z angular power spectrum in Fig.~\ref{fig:Ycl_evo_comp}.
Around the peak,
the main contribution comes from massive halos ($M>10^{15} M_\odot$).
However, halos with masses smaller than $10^{15} M_\odot$ become important
in the high $\ell$ modes.
The contribution from halos with masses $10^{14}M_\odot <M <10^{15}M_\odot$
and with masses $M<10^{14}M_\odot$ dominate
at $\ell \sim 10000$ and $\ell >50000$, respectively.
For comparison, we plot the spectrum of the S-Z effect without gas condensation
as a thin line.
The smaller the halos are, the more efficient  gas condensation
becomes.
Therefore, the difference between the spectra with and
without gas condensation is large for the high $\ell$ modes.

\begin{figure}
  \begin{tabular}{cc}
   \begin{minipage}{0.5\textwidth}
  \begin{center}
    \includegraphics[keepaspectratio=true,height=50mm]{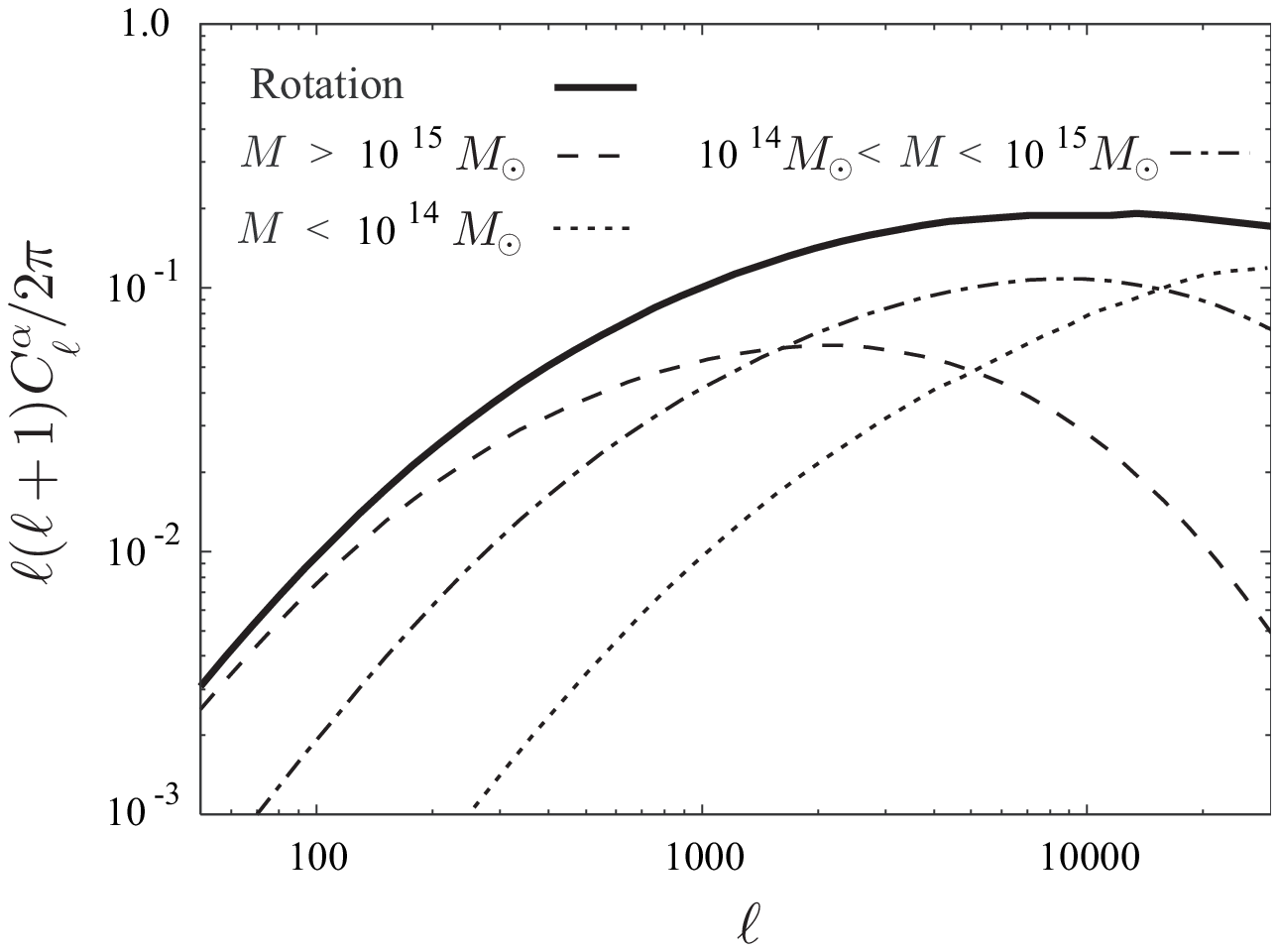}
  \end{center}
  \end{minipage}
   \begin{minipage}{0.5\textwidth}
  \begin{center}
    \includegraphics[keepaspectratio=true,height=50mm]{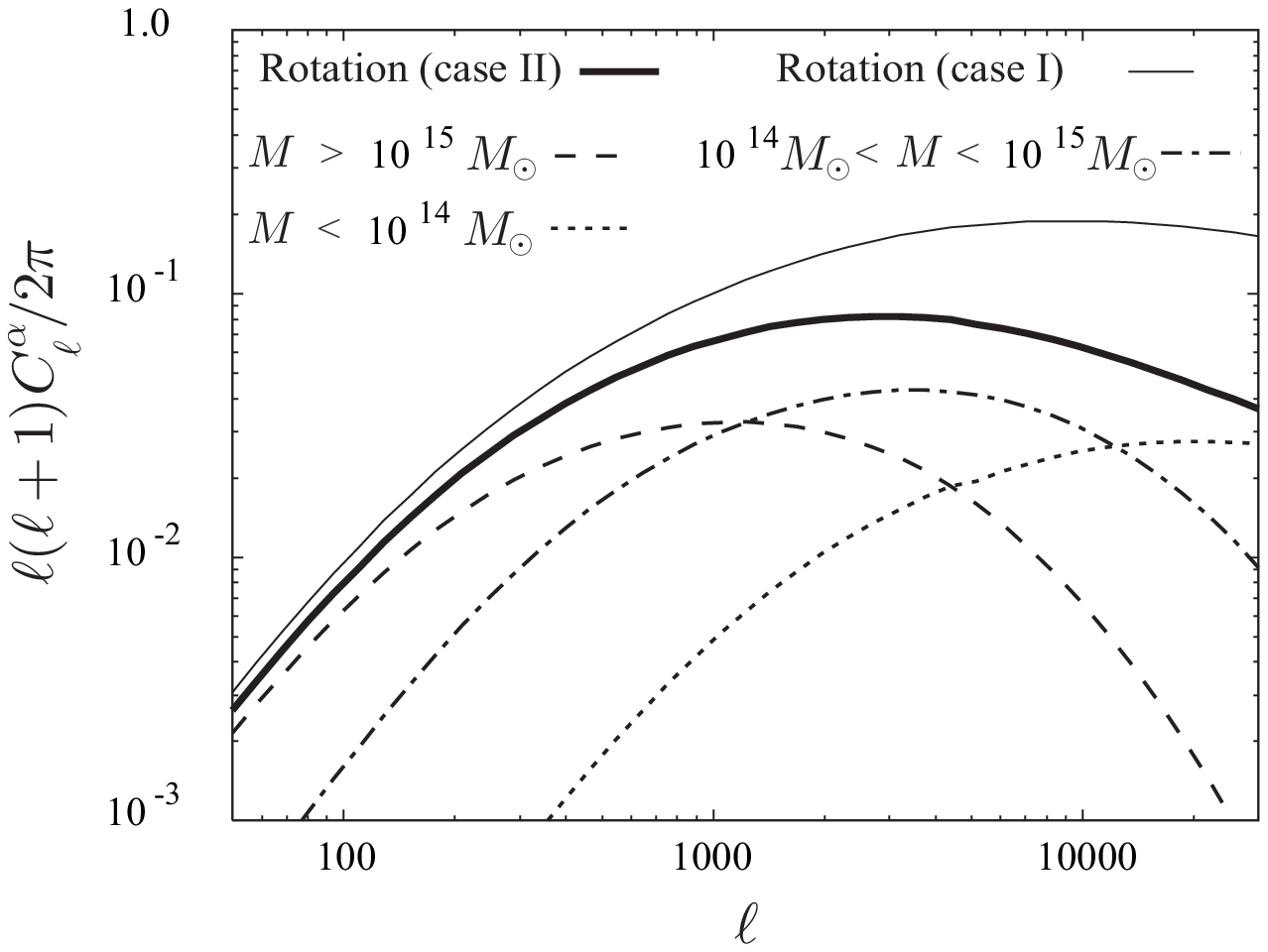}
  \end{center}
   \end{minipage}
  \end{tabular}
  \caption{Angular power spectra of rotation measurements.
  The contributions from halos with $M < 10^{14} M_\odot$, 
  $10^{14} M_\odot <M < 10^{15} M_\odot$ and $M> 10^{15} M_\odot$ are given 
  as the dotted, dotted-dashed, dashed lines, respectively.
  We set the CMB frequency to 30 GHz.
  In the left panel, we plot the angular power spectra in case I. 
  In the right panel, we show the spectra in  case II. For comparison, we plot the spectrum
  in  case I as a thin solid line.}
  \label{fig:ROTcl_evo}
\end{figure}

Next, we show the angular power spectrum of Faraday rotation measurements
in Fig.~\ref{fig:ROTcl_evo}.
In this figure, we set the CMB frequency to $30$ GHz.
The position of the peak is around $\ell = 5000$ or $\ell=10000$,
depending on the magnetic field evolution.
This peak position is on higher $\ell$ than for the S-Z power spectrum and
the main contribution around the peak is produced by halos with 
$10^{14}M_\odot <M <10^{15}M_\odot$.
Compared with the S-Z effect,
halos with small masses are more important in 
rotation measurements.
The mass dependence of the S-Z effect 
is $y \sim n_e T_e R \propto M$, while
that of rotation measurements is $RM \sim n_e B R \propto M^{2/3}$.
This fact makes the scale where the S-Z contribution from halos with small masses 
dominate shift to low $\ell$ modes.
For rotation measurements, 
halos with $M >10^{15} M_\odot$ 
dominate around $\ell \sim 1000$ and
halos with mass $10^{14}M_\odot <M <10^{15}M_\odot$
and with mass $M<10^{14}M_\odot$ overwhelm other components
at $\ell \sim 6000$ and $\ell >20000$, respectively.
In the left panel of Fig.~\ref{fig:ROTcl_evo},
we show the dependence on the magnetic field evolution.
Rapid evolution as in  case II suppresses the amplitude 
at high $\ell$ modes ($\ell > 2000$),
because these are produced by halos around $z \sim 0.5$
where the magnetic field energy becomes  half of its present-day value
(see also Fig.~6 in \citealt{tashiro-faraday}).

\begin{figure}
  \begin{tabular}{cc}
   \begin{minipage}{0.5\textwidth}
  \begin{center}
    \includegraphics[keepaspectratio=true,height=50mm]{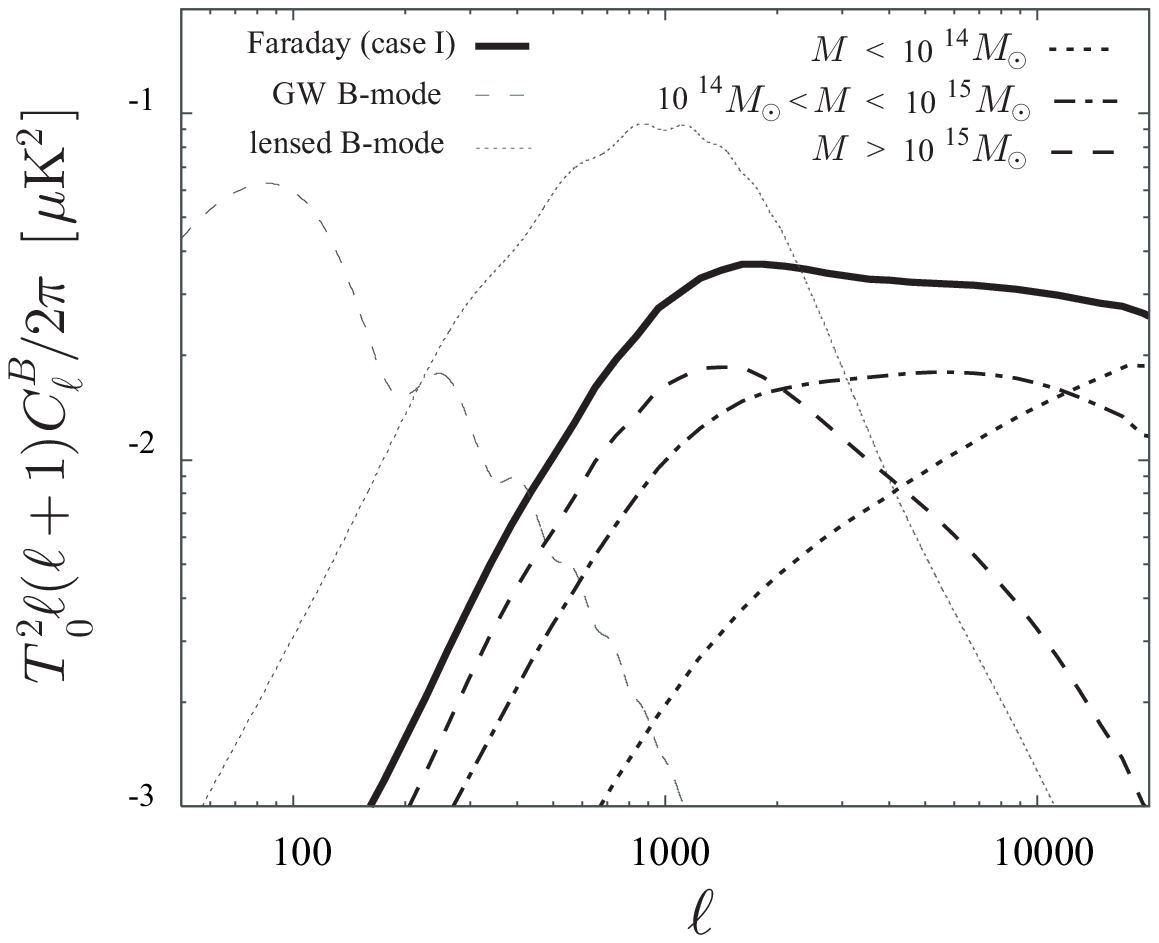}
  \end{center}
  \end{minipage}
   \begin{minipage}{0.5\textwidth}
  \begin{center}
    \includegraphics[keepaspectratio=true,height=50mm]{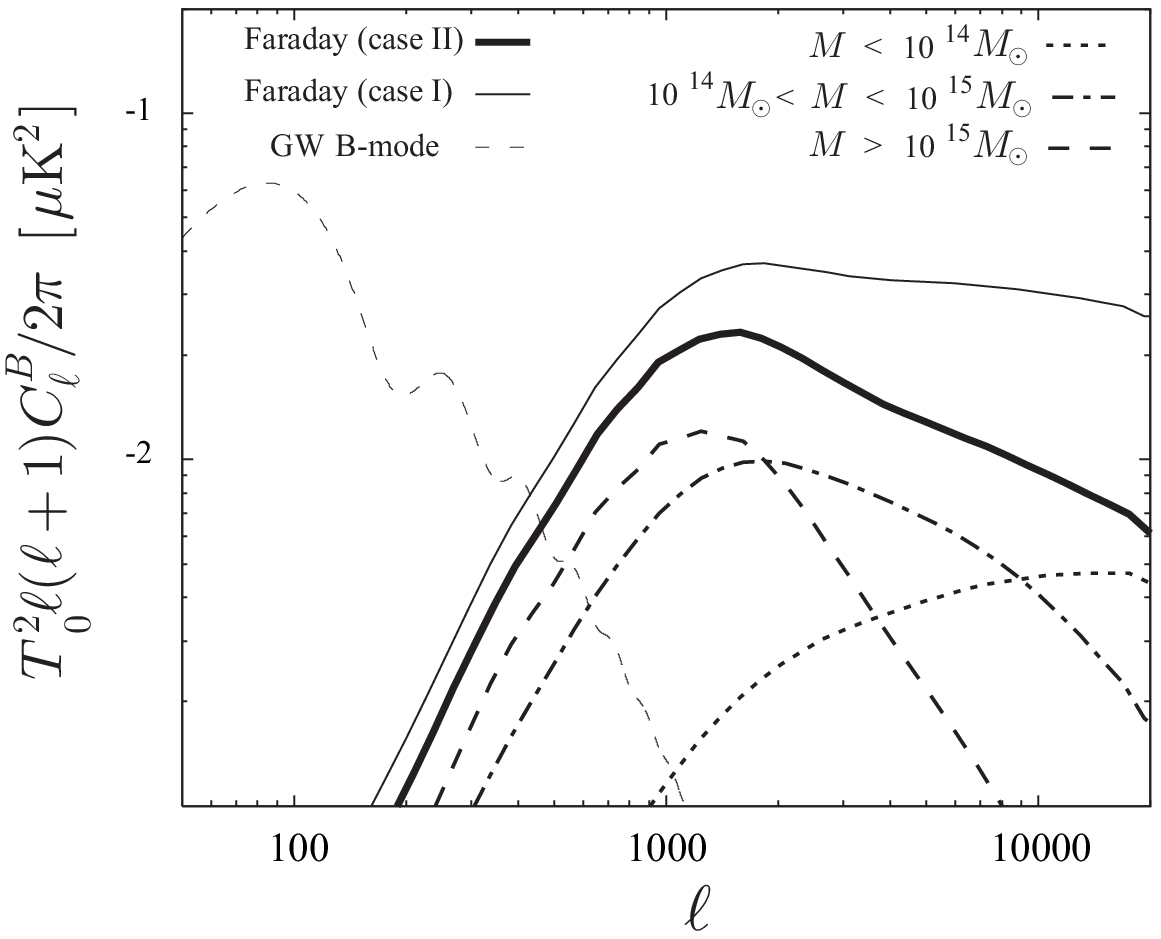}
  \end{center}
   \end{minipage}
  \end{tabular}
  \caption{Angular power spectra of CMB $B$-mode polarisation produced by Faraday rotation.
  In both panels, the dotted, dotted-dashed, dashed lines represent
  the contributions from halos with $M < 10^{14} M_\odot$, 
  $10^{14} M_\odot <M < 10^{15} M_\odot$ and $M> 10^{15} M_\odot$, respectively.
  We set the CMB frequency to 30 GHz.
  In the left panel, we plot the $B$-mode angular power spectra in  case I 
  of the magnetic field evolution. For comparison, we put the GW $B$-mode and lensed $B$-mode
  as the thin dashed and thin dotted lines, respectively.
  In the right panel, we show the spectra in case II. For comparison, we give the $B$-mode spectrum
  in case I as the thin solid line.}
  \label{fig:FARAcl_evo}
\end{figure}

The angular power spectra of CMB $B$-mode polarisation induced by
Faraday rotation is shown in Fig.~\ref{fig:FARAcl_evo}.
These power spectra reflect the angular power spectra of the rotation measurements.
The effect of small halos arises at lower $\ell$ modes 
than for the S-Z effect.
Although the contribution from halos with $M < 10^{15} M_\odot$
is sub-dominant for the S-Z power spectrum,
these still provide a major contribution to the Faraday $B$-mode spectrum.
Halos with $10^{14} M_\odot< M < 10^{15} M_\odot$
dominate over halos with larger masses for $\ell > 2000$.
This should have  important observational implications.  
On scales where the contributions of halos with $M> 10^{15} M_\odot$
are dominant, the $B$-modes from Faraday rotation are overwhelmed by the $B$-modes 
from gravitational lensing. 
The detection of the $B$-modes from Faraday rotation is expected 
for $\ell$ modes bigger than $\ell = 3000$.
The gas and magnetic field properties of galaxy groups
and galaxies are important for  the $B$-modes produced by Faraday rotation.

Regarding magnetic field evolution, 
as shown in the left panel of Fig.~\ref{fig:FARAcl_evo},
an effect is apparent on small scales.
Rapid magnetic field evolution as in  case II strongly
suppresses the amplitude of high $\ell$ modes.
This is also related to the decrease of the Faraday rotation
in small mass halos. 
We conclude that the measurement of $B$-mode Faraday rotation can
put constraints on gas and magnetic field evolution.

\section{conclusion}

In this paper, we examine the S-Z effect and Faraday rotation
from halos which have a wide mass range ($10^{12}M_\odot < M <10^{17}M_\odot$).
In particular, we take into account the effects of gas cooling 
based on recent observations
and consider two cases for magnetic field evolution
motivated by theoretical considerations
and numerical simulations.

First, we calculate the halo number counts for the S-Z effect.
We find that the number counts for the S-Z effect are sensitive to gas condensation.
The redshift distribution is suppressed by gas condensation over all redshifts. 
In the mass distribution, gas condensation 
suppresses the number counts on  small scales.
As a result, 
the peak position of the distributions
shifts to larger mass scales.

We also calculate the rotation measurement for S-Z halos.
The distribution of number counts strongly depends
on the magnetic field evolution.
The detectable minimum mass is decided by the threshold of rotation measure.
Therefore, by varying the threshold 
we can obtain number counts of S-Z halos over  the
corresponding mass scales and  constrain the magnetic field evolution.

We study the CMB angular power spectrum 
arising from the S-Z effect and Faraday rotation in galaxy clusters, galaxy groups 
and galactic halos.
For the S-Z effect, the main contribution comes from massive halos ($M>10^{15} M_\odot$).
These set  the peak at around $\ell \sim 2000$.
Halos with mass $M<10^{14}M_\odot$
dominate the contribution on small scales.
The S-Z effect in halos with mass $10^{14}M_\odot <M <10^{15}M_\odot$
and with mass $M<10^{14}M_\odot$ 
produces CMB anisotropies
over $\ell \sim 10000$ and $\ell >50000$, respectively.
Gas condensation is effective on small mass scales
so that it suppresses the CMB anisotropies at $\ell \sim 10000$.

For the $B$-mode polarisation, 
small mass halos play more important roles than 
for the S-Z CMB anisotropies.
As a result, the $B$-modes produced in such small halos 
arise on larger $\ell$ scales than the S-Z CMB anisotropies.
The peak of the $B$-mode power spectrum
is produced by
halos with $10^{14} M_\odot< M < 10^{15} M_\odot$
and  is located at   $1000 < \ell <5000$.
Halos with $M >10^{15} M_\odot$ 
are dominant around $\ell \sim 1000$ and
$B$-modes produced by halos with mass $M<10^{14}M_\odot$ 
arise at  $\ell >20000$, respectively.
The evolution of magnetic fields modifies the amplitude
produced by halos with $M <10^{14} M_\odot$.
The small mass halos at  high redshift  can
collectively generate detectable  $B$-mode polarisation.
However, 
since the halos at high redshift 
are easily affected  by the evolution of
magnetic fields, the $B$-modes by such halos
are rapidly damped in the rapid evolution case.  

Comparing to other $B$-mode polarisation contributions,
the $B$-modes from Faraday rotation
dominate the secondary $B$-modes caused by gravitational
lensing at $\ell < 3000$.  
On these scales, the $B$-modes by Faraday rotation
are produced by halos with mass $M<10^{14}M_\odot$.
Therefore, measurement of such $B$-mode polarisation
could  put constraints on magnetic field evolution,
in combination with the S-Z power spectrum.

\section*{Acknowledgements}
NS is supported by a Grant-in-Aid for Scientific Research 
from the Japanese Ministry of Education (No. 17540276).

\end{document}